\documentclass[9pt,twocolumn,preprintnumbers,amsmath,amssymb,superscriptaddress,floatfix]{revtex4}

\usepackage{latexsym}
\usepackage{epsfig}
\usepackage{graphicx}
\usepackage{textcomp}
\usepackage{color}
\usepackage{amsmath}

\begin{document}
\title{Pulses from a mid-infrared quantum cascade laser frequency comb using an external compressor}
\author{Matthew Singleton}
\email{smatthew@phys.ethz.ch}
\affiliation{Institute for Quantum Electronics, ETH Zurich, 8093 Z\"urich, Switzerland}
\author{Mattias Beck}
\affiliation{Institute for Quantum Electronics, ETH Zurich, 8093 Z\"urich, Switzerland}
\author{J\'er\^ome Faist}
\affiliation{Institute for Quantum Electronics, ETH Zurich, 8093 Z\"urich, Switzerland}

\begin{abstract}
A Martinez-type stretcher-compressor is used to modify the spectral phases of a high-power ($\sim$1~W) QCL comb emitted at 8.2~ \textmu m with more than 100 cm$^{-1}$ spectral bandwidth.
Using this scheme, we demonstrate a compression of the QCL output from a 134~ps continuous wave waveform, to a train of pulses of width 12~ps, and a power with peak to average ratio of 40.7.
An evaluation of the phase noise of the free-running device yields an integrated timing jitter of 335~fs over the frequency range 20~kHz - 100~MHz, and a pulse-to-pulse jitter of 2.0~fs.
\end{abstract}
\maketitle
\newcommand{\lasercurrent}{1.1688~A}

\label{sec:intro}
Ultrashort physics in the mid-infrared is an underdeveloped field with exciting prospects\cite{pires_ultrashort_2015}.
High peak power sources at these wavelengths find application in nonlinear spectroscopy\cite{meek_doppler-free_2018,sugiharto_generation_2008}, supercontinuum generation\cite{kubat_mid-infrared_2014}, and standoff detection\cite{katz_standoff_2008,ting_remote_2005}, for example.
For more energetic mid-IR pulses, favourable wavelength scaling of electron rescattering processes make these interesting sources for soft x-ray generation and related applications\cite{yakovlev_enhanced_2007,cousin_high-flux_2014,colosimo_scaling_2008,blaga_imaging_2012}.
A number of approaches, chiefly based on difference frequency generation\cite{kaindl_broadband_1999,sugiharto_generation_2008,krogen_generation_2017,phillips_widely_2012,golubovic_all-solid-state_1998}, and cascaded optical parametric amplifiers and oscillators\cite{mcgowan_femtosecond_1997,hemmer_18-j_2013,liang_high-energy_2017,mayer_sub-four-cycle_2013,pupeza_high-power_2015,kanai_parametric_2017} have been used to generate pulses in the mid-infrared, with microjoule sub two optical cycle pulses having been demonstrated\cite{chaitanya_kumar_few-cycle_2014}.
The common element to these solutions is that they are tabletop scale;
to be able to directly generate mid-infrared pulses in a monolithic platform could significantly broaden the applicability of such systems, facilitating the study of new physics.

As proven stable\cite{cappelli_frequency_2016,bartalini_direct_2018}, compact, monolithic comb sources\cite{hugi_mid-infrared_2012,burghoff_terahertz_2014}, quantum cascade lasers (QCLs) would make for an interesting system to produce pulses in the mid-infrared and terahertz.
However, the very mechanism responsible for the phase locking, the short sub-ps upper state lifetime\cite{choi_gain_2008} which drives the broadband four-wave mixing, is that which makes it extremely difficult to internally form singular, isolated pulses intracavity~\cite{kuehn_ultrafast_2008,choi_gain_2008}.
Direct phase measurements confirm the laser output to exhibit a continuous wave character with a strong frequency modulation, where the laser periodically sweeps its bandwidth\cite{singleton_evidence_2018};
this well supports previous conclusions drawn from indirect measurements\cite{hugi_mid-infrared_2012, khurgin_coherent_2014}.

Nonetheless, some techniques have succeeded in producing pulses in QCLs.
A QCL lasing at 6.3~\textmu m, with an active region specially engineered for a long gain recovery time ($\sim$50 ps), and an electrically isolated short section, used to modulate the loss at the cavity roundtrip time, was able to produce transform limited pulses on the order of 3~ps\cite{wang_mode-locked_2009}.
However, the device had to be operated at cryogenic temperatures (77 K) close to the lasing threshold, limiting the fractional bandwidth to about 1.5\% at 1585~cm$^{-1}$, and the peak power to less than 10 mW (0.5~pJ per pulse).
In the terahertz, the situation is somewhat more favourable, with the gain recovery time naturally in the 10s of picoseconds~\cite{green_gain_2009,bacon_gain_2016}, comparable to a short cavity roundtrip time. 
One approach taken by Oustinov and coworkers\cite{oustinov_phase_2010} was based on synchronising the RF modulation to a THz seed pulse injected directly into the cavity, made possible by the closely matched effective indices for the RF mode and the terahertz mode.
In this way, they were able to amplify these naturally transform-limited pulses over several round-trips, with pulse widths measured at around 11~ps FWHM.

An alternative approach to direct pulse generation is to compensate for the intermodal phase differences in the laser output field externally.
This has already been performed for microresonator combs at 1550~nm using the technique line-by-line pulse shaping\cite{weiner_femtosecond_1990,weiner_programmable_1990,weiner_femtosecond_1995,jiang_spectral_2005}, where liquid-crystal waveshapers modify the phase of individual spectral components.
An optimisation routine stepwise changes the applied phase with a view to maximising the second-harmonic signal, thereby ultimately to producing near transform-limited pulses, and characterising the natural comb state\cite{ferdous_spectral_2011,delhaye_phase_2015}.
External pulse compression is advantageous in that the device can be operated under normal conditions, hence making the most effective use of the gain medium. As such, the bandwidth will be naturally wide, and the average power in the Watt range, which, in the case of a perfect phase correction, would translate to pulses of widths on the order of 300~fs and peak powers of several hundred Watts.
Certain properties of the QCL comb states, namely their stability, repeatability, and mainly quadratic phase profile\cite{bartalini_direct_2018,singleton_evidence_2018}, lend themselves to a fixed and potentially simple setup to correct the phase excursions.
In this work we demonstrate the phase compensation to the first order using a grating compressor.

\begin{center}
	\begin{figure}
		\includegraphics[width=\linewidth]{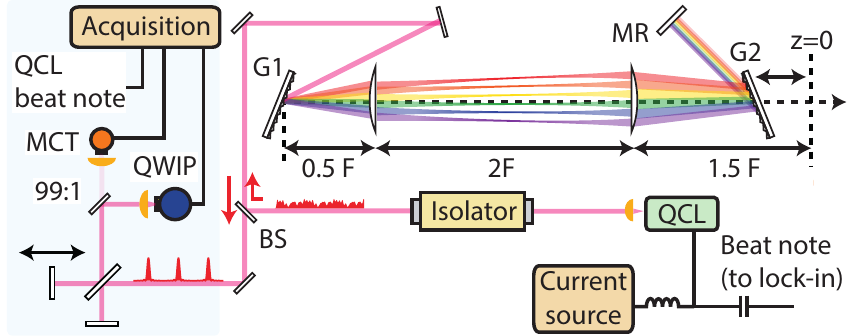}
		\caption{\label{fig:setup}
			Experimental setup. QWIP (Quantum Well Infrared Photodetector); MCT (Mercury Cadmium Telluride detector);
			A pair of matched ruled diffraction gratings (G1, G2), a pair of lenses (F=10~cm), and the retroreflecting mirror MR form the compressor. G2 and MR are mounted on a common translation stage, directed along $z$. The optical axis $z$ is positive in the right hand direction, and so a displacement of G2 left indicates a positive GDD, and vice-versa.
		}
	\end{figure}
\end{center}

\section{Experiment}
Our sample is a 6~mm mid-IR QCL lasing at 8.2~\textmu m, featuring a plasmon-enhanced waveguide for dispersion compensation\cite{jouy_dual_2017}.
The device operates as a comb, lasing on hundreds of modes and showing a singular sharp peak in the radio frequency spectrum at the inverse cavity round-trip time. 
This device was previously shown to exhibit the reproducible linear chirped state\cite{singleton_evidence_2018}, where the spectral phases follow a mainly quadratic profile, quantified by the field group delay dispersion (GDD):

\begin{equation}
\label{eq:gdd}
GDD = \frac{\partial^2 \phi}{\partial \omega^2}
\end{equation}

We run the device at 291~K, with a DC bias of \lasercurrent, where we find the laser output field to have a GDD of $-6.45$~ps$^2$ (see Fig.~\ref{fig:pulses}~(b) at 0~cm, or Fig.~\ref{fig:afig_device_char}).

Our experimental setup shown in Fig.~\ref{fig:setup} comprises the QCL, a compressor to compensate the second order dispersion, and a Fourier transform spectroscopy measurement based on SWIFTS (Shifted Wave Interference Fourier Transform Spectroscopy)\cite{burghoff_terahertz_2014, burghoff_evaluating_2015} (see Appendix~A).
The latter is a radio-frequency (RF) technique which gives direct access to the intermodal phase differences, and has been used to characterise the temporal behaviour of QCLs\cite{burghoff_evaluating_2015,singleton_evidence_2018}.
The QCL light is collimated using a high numerical aperture collimation lens (focal length 1.173~mm) and passed through an optical isolator to mitigate feedback.
A beamsplitter then directs the beam to the compressor, which modifies only the spectral phases, and reflects the light directly back to the beamsplitter.
The modified light is then sent through a Michelson interferometer and shined onto a sensitive MCT and fast QWIP (splitting ratio 99:1), meaning the DC autocorrelation trace and phase information containing correlation (SWIFTS) trace can be recorded simultaneously.

The stretcher-compressor is the Martinez type\cite{martinez_3000_1987}, made up of a pair of antiparallel gratings (G1, G2: 150 lines/mm, blazed 35$^\circ$) and a pair of identical achromatic doublet lenses (focal length F=10~cm) separated by twice the focal length to form a relay pair.
A mirror retroreflects the light directly back through the system, which both doubles the effective dispersion induced on the field, and reverses the the spatial chirp.
The second grating G2 and retroreflecting mirror MR are mounted on a motorised translation stage, allowing their position to be adjusted along the optical axis $z$, while keeping their relative positions fixed.
Note that the relay pair are shifted 5~cm towards the first grating, extending the travel of the second grating.
The effective group delay dispersion per unit displacement is set by the angle of incidence for a fixed central wavelength, which for our system is about 54.5 degrees ($\sim 0.53$~ps$^2$/cm).
A larger dispersion would be obtained for a smaller incident angle, but at the price of a more difficult geometry and, in the absence of larger optics, spectral clipping.
At the origin $z=0$, where the two gratings are separated by exactly 4F, the system is nondispersive, and the field is returned simply attenuated.
By moving the second grating towards the first grating ($z<0$), a positive GDD is added corresponding to 0.53~ps$^2$ per cm displacement.

The measurement proceeds as follows. The delay stage is set to around $-14$~cm, which is the farthest possible position without clipping of the beam.
It is then stepped away in 2~mm increments such that the separation between the two gratings increases.
At each grating displacement along $z$, the standard autocorrelation and SWIFTS traces are recorded (resolution 0.12~cm$^{-1}$), giving complete spectral amplitude $A_n(z)$ and phase information $\phi_n(z)$ for each comb mode $n$.
We can therefore reconstruct the instantaneous intensity by the inverse Fourier transform:

\begin{equation}
\label{eq:intensity}
I(t,z) = \left | \sum_n A_n e^{i(\omega_0 + n\omega_r)t + \phi_n(z)} \right |^2
\end{equation}

$\omega_0$ and  $\omega_r$ are the usual carrier envelope offset frequency and comb repetition rate, in rad/s, and $A_n(z)\approx A_n$ but for the effect of finite beam size\cite{martinez_grating_1986}, aberrations, and absorption.

\section{Results}
\begin{figure*}[!htb]
	\centering
	\includegraphics[]{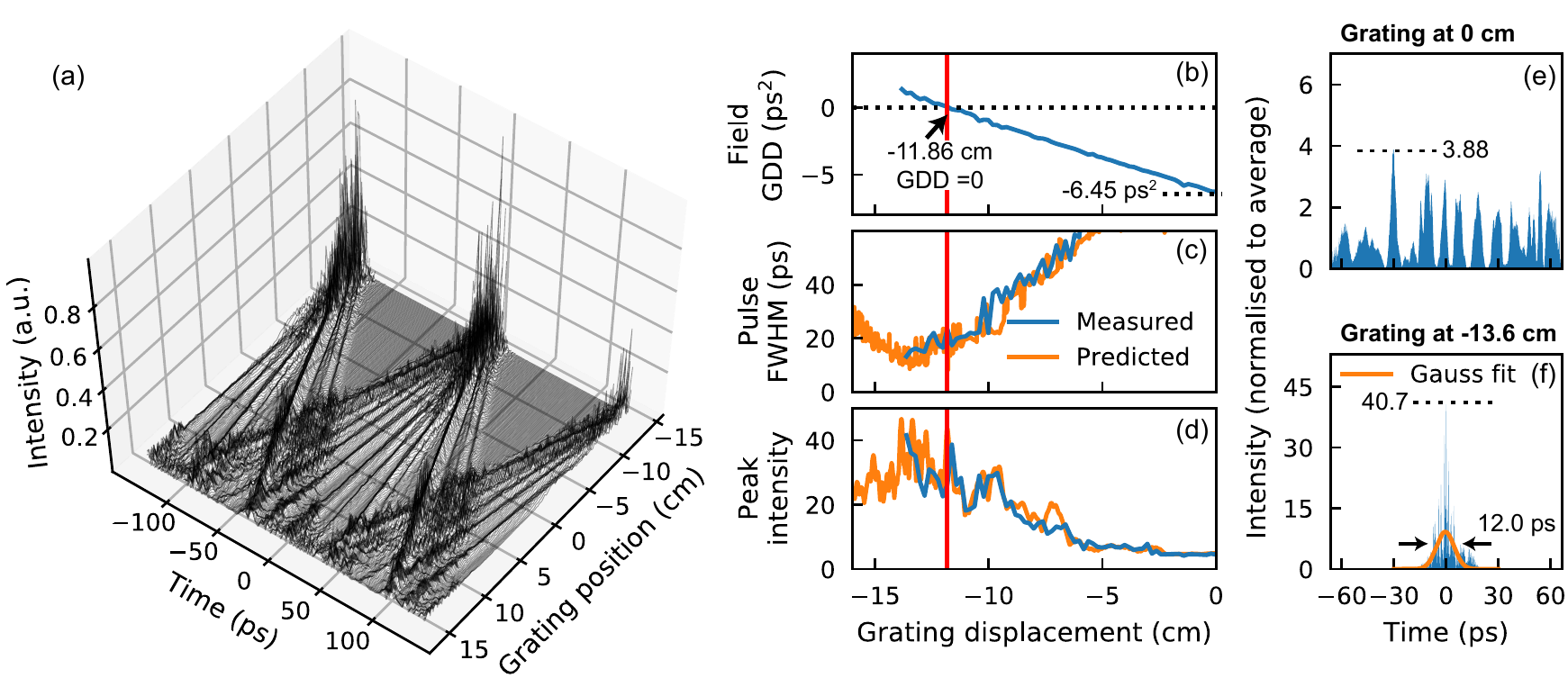} 
	\caption{\label{fig:pulses}  
		(a) Instantaneous intensity as a function of the grating position, calculated from the measured complex spectra by Eqn.~\ref{eq:intensity}.
		As the grating G2 approaches in the direction of G1, indicated by the negative displacement, the pulse narrows and the peak power increases. 
		(b) The field GDD, calculated from fits to each of the measured complex spectra.
		The red line at -11.86~cm indicates the position at which the quadratic part of the phase has been completely eliminated.
		-6.45~ps$^2$ marks the dispersion at $z=0$.
		(c) Pulse width (Gaussian FWHM) and (d) peak intensity, normalised to the average, as a function of grating position.
		The orange curves are calculated by applying the relevant phase shifts given by the grating pair to the reference measurement at $z=0$.
		(e) Comparison of the waveform at zero displacement and (f) position with highest peak power.}
\end{figure*}
In Fig.~\ref{fig:pulses}~(a) we show how the steady-state instantaneous intensity evolves as a function of the grating displacement (Eqn.~\ref{eq:intensity}).
It can be seen that that, as the displacement becomes more negative, the energy becomes more compressed within the period and a pulse is shown to emerge from the background.
Conversely when the grating moves in the positive direction, interferences between the adjacent comb periods can be seen in the form of high frequency oscillations.

The field group delay dispersion, estimated by a fit to the measured phase traces at each position, is shown in (b) to change linearly with the grating position, as expected.
We also show in (c) the (fitted) Gaussian equivalent pulse width and in (d) the peak intensity evolving with the grating position, both compared to the values predicted when applying the expected phase correction to the phase spectrum measured at 0~cm.
The calculated points were deliberately extended past the geometric constraint of the setup to highlight that this is not the limiting factor, with the performance in terms of pulse width and peak power expected to quickly degrade past -14~cm.
The red line in the (b-d) indicates the position at which the quadratic part of the phase has gone to zero.
Note that this coincides with neither the maximum peak power, nor the shortest pulse.
The discrepancy arises predominantly due to the stronger negative dispersion in the redmost lobe of the spectrum, which contains a substantial part of the power;
the GDD evaluated at a central wavelength is an overly simplistic metric of the spectral phase profile for these lasers, especially in the context of compression (see Fig.~\ref{fig:afig_device_char}). 

Cuts of the instantaneous intensity at the $z=0$ and $z=-13.6$~cm are shown in Fig.~\ref{fig:pulses} (e) and (f), respectively.
A pulse of equivalent Gaussian FWHM 12~ps is observed to emerge, with a peak to average ratio of 40.7.
In the ideal, transform limited case, with the DC spectrum as is and the phases all set identical, a pulse with peak to average ratio in excess of 320 would be expected, with a pulse width less than 300~fs.
At an estimated optical bandwidth of 2.79~THz at 10~dB, the Fourier limited pulse exhibits a time-bandwidth product of 0.8, limited by the rectangular spectral shape which gives rise to a quasi-sinc pulse.
An equivalent Gaussian spectrum would give a pulse length of $\sim 0.44/2.79$~THz~$= 157$~fs, though to shape the full amplitude spectrum would strongly reduce the integrated pulse energy.

Though we start with 600~mW, accounting for the loss in the final beamsplitter, only about 14~mW is observed to emerge from the system.
This means our pulse has a peak power $P_0 = P_{avg} I_0/I_{avg}$ of about 574 mW, with a pulse energy $P_{avg}/f_{rep}\approx E$ 1.9~pJ.
52\% of the intensity is lost at the optical isolator, and the rest can be accounted for through the beamsplitter, the non-perpendicularity of the incident light with respect to the grating rulings, and lossy coatings present on some of the mirrors.

\begin{figure}[!ht]
	\centering
	\includegraphics[]{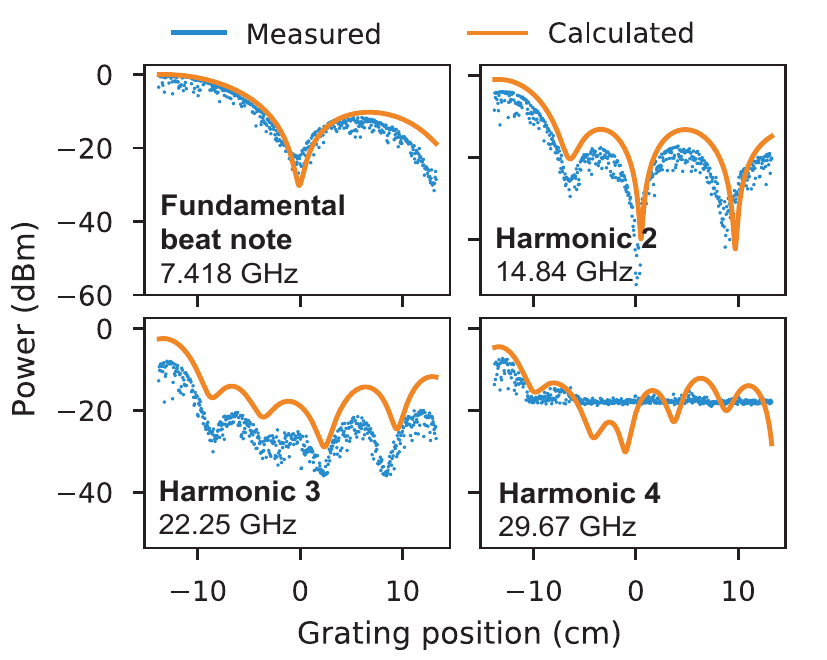} 
	\caption{\label{fig:beatnote_traces}Beat note power at the fundamental, second, third, and fourth electronic harmonics measured as a function of grating position, with the calculated traces superimposed (orange lines).
	}
\end{figure}
In addition to the fundamental beat note, which we use to characterise the phase spectrum, we also detect the second, third, and fourth radio frequency harmonics on the QWIP.
In Fig.~\ref{fig:beatnote_traces} we measure the power in the first four harmonics using a spectrum analyser (Rohde \& Schwarz FSW) as a function of grating displacement, and compare their evolution to that predicted.
For each beating order $b$, the expected power is given by:
\begin{equation}
B_b(z)=\left |\sum_n A_n A_{n-b}^*e^{i(\phi_C(\omega_n, z)-\phi_C(\omega_{n-b}, z))}\right|^2 \left | \tilde{H}(b\omega_r) \right |
\label{eq:beatnote}
\end{equation}

Here, the complex modal amplitudes $A_n=|A_n|e^{i\phi_n}$ are taken from the reference measurement at $z=0$ only, and $\phi_C(\omega,z)$ denotes the phase compensation at $\omega$ imparted by the grating pair when set to position $z$.
We also account for the combined magnitude response of the QWIP itself and components which come between it and the spectrum analyser through $|\tilde{H}(\omega)|$, which we estimate using a rectification technique\cite{liu_high-frequency_1996} under the same conditions we operate the detector during the measurement (77~K, biased at 4~V).

As expected for an FM comb\cite{hugi_mid-infrared_2012,khurgin_coherent_2014}, the beatings are shown to be strongly suppressed at $z=0$, increasing as a general trend with negative grating displacement.
Strikingly, at the position of maximum peak intensity (-13.6~cm), the beatings are shown to be simultaneously maximised, a key signature for an AM pulse train where all the phases have the same sign and similar magnitude.
These behaviours are reflected in the calculated beat note traces, which show good qualitative agreement for all measured grating displacements.
\begin{figure*}[!htb]%
	\centering
	\includegraphics[]{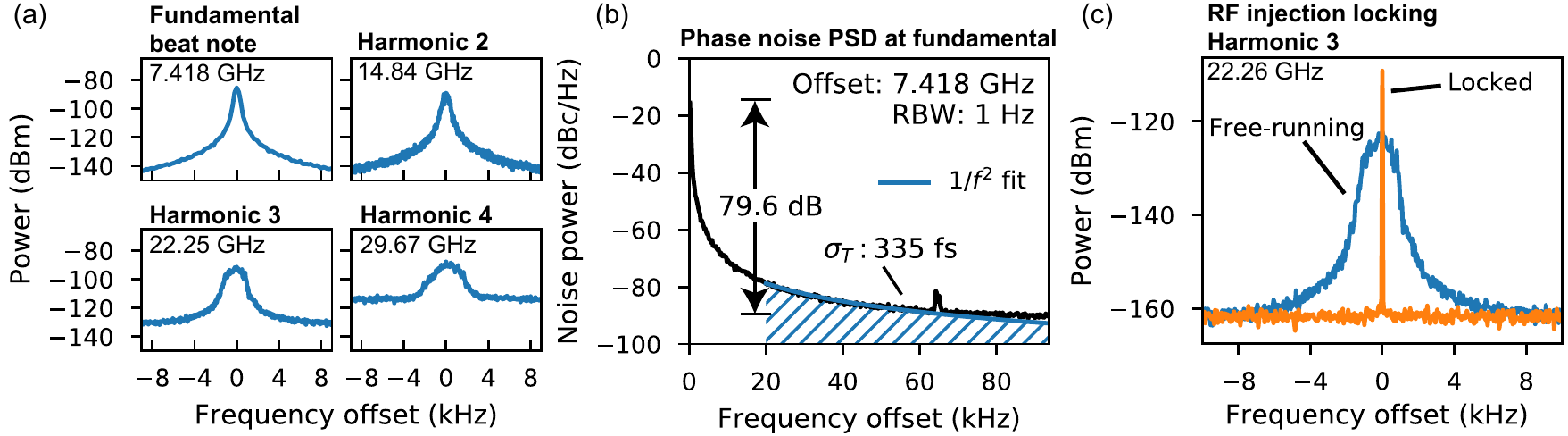} 
	\caption{\label{fig:jitter}
		Radio frequency spectra as recorded on a fast detector (QWIP) with a spectrum analyser, with the grating at the position where the field shows the strongest pulses.
		The resolution bandwidth (RBW) is 1 Hz for all measurements
		(a) Comparison of the first four harmonics, showing the expected broadening.
		(b) Radio frequency spectrum of the fundamental beat note as measured at a 1~Hz resolution bandwidth (RBW), scaled to the total power in the carrier, evaluated at a 200~Hz RBW.
		The blue line is a $1/f^2$ fit to the tail, from which the integrated timing jitter is inferred.
		(c) Third harmonic electronic beat note, both free-running and when the laser has been RF injection locked at the fundamental beat frequency (7.42~GHz).}
\end{figure*}
Traditionally, the fast noise dynamics of mode-locked lasers was characterised by the timing jitter of the pulse train.
Indeed, the timing jitter, described either in terms of its integrated value $\sigma_T$ or its pulse-to-pulse value $\sigma_{pp}$, can be conveniently extracted from an integral of the electrical noise power spectral density on a fast detector\cite{von_der_linde_characterization_1986}, without relying on nonlinear detection methods (e.g., \cite{jiang_measuring_2002}).
This technique is especially well-suited for the characterisation of mode-locked lasers with cavities in the metre range, such that the optical modes are tightly spaced in the frequency domain and the electronic beat notes easily accessible, being typically in the MHz.
While the approach is most appropriate for actively mode-locked lasers, with some care similar inferences can be made about passively mode locked lasers such as ours\cite{kefelian_rf_2008,eliyahu_noise_1996}.
In contrast with the normal QCL emission with its strong FM characteristics, the pulsed mode-locked emission of our device after correction enables a straightforward application of Von der Linde's analysis\cite{von_der_linde_characterization_1986}. 

Shown in Fig.~\ref{fig:jitter} is the power spectral density of the beat note at the fundamental round trip frequency compared to the power of the carrier at the centre of the band (dBc), showing a dynamical range of almost 80~dB at the point of maximum compression.
We find an integrated timing jitter 20~kHz~-~100 MHz of 335~fs, and a pulse-to-pulse timing jitter of 2.0~fs, with reasonable agreement between the different orders (see Appendix~D).
The same analysis applied to the measured beat note of a QCL whose FM output was converted to an AM one using an optical discriminator yielded an integrated timing jitter of 342~fs, which is very close to that found for our device. 
In general, our values compare well with those measured on quantum dot semiconductor passively mode-locked lasers, which typically feature broader beat notes on the order of several kHz.
As an example, \cite{kefelian_rf_2008} report an integrated jitter value of 1.6~ps for such a device over the same frequency range of 20~kHz to 100~MHz. 

Jitter performance can be drastically improved by injecting RF from a stable source at approximately the fundamental beat frequency, which acts to lock the comb repetition rate.
QCLs have an intrinsically high electronic cut-off frequency, owing to the lack of relaxation peak\cite{paiella_high-frequency_2001}, and remain responsive to bias modulations above 10 GHz\cite{hinkov_rf-modulation_2016};
this makes them particularly amenable candidates for RF injection locking, and this has been successfully demonstrated\cite{ravaro_stabilization_2013,st-jean_injection_2014}.
Importantly, all lasing modes can be coherently locked to the external oscillator, and this has been shown not to perturb the comb state\cite{hillbrand_coherent_2019}.
We show in Fig.~\ref{fig:jitter} the dramatic difference between the 3rd order beat note for the free running QCL, and that observed when the QCL has been injected with RF at the fundamental of around 7.42~GHz.

\section{Conclusion}
We have demonstrated the compression of a free-running continuous wave QCL comb, of period $\sim134$~ps, into a train of well-defined pulses of around 12~ps FWHM.
With this pulse train, we were able to characterise the timing jitter of our comb.
That we are able to achieve this correction with a grating compressor, a staple of the ultrafast community, gives an additional strong confirmation of the intrinsic stability and reproducibility of the intermodal phases.

While the peak to average ratio increases to 40.7 at the optimal location, a tenfold increase from the uncompressed output, losses in the experiment mean that there is no overall increase in peak power.
Some of the responsible issues can be mitigated through optimisation of the setup, but we are fundamentally limited by the higher order dispersion present in the field.
To compensate for this, a more sophisticated scheme would have to be used.
For example, prisms could be placed in the focal plane between the two lenses: this would be of most use on the wings of the spectrum, which have a stronger dispersion quite dissimilar from the central portion.
One could also take advantage of the various aberrations to tailor the dispersion up to the fourth order\cite{white_compensation_1993}, at the price of the greater complexity that comes with the additional degrees of freedom.
In the other direction, as the spectrum is deterministic, a fixed phase mask in the Fourier plane, or indeed a tunable thermal slab\cite{osvay_high_2007}, could in principle compensate the dispersion almost entirely for a given state.

This proof-of-concept experiment demonstrates that external compensation schemes are a feasible approach to controlling the phases of the QCL comb.
A spatial light modulator, for example, could be used to not only negate the field dispersion, but would allow for total control of the individual amplitudes and phases\cite{weiner_programmable_1990}, enabling arbitrary tailoring of the QCL waveform, limited only by the intrinsic period, bandwidth, and power of the laser.
In the context of compression, it should be possible to convert the typical QCL comb emission into a train of stable pulses with favourable time-bandwidth products, and potentially peak powers in the hundreds of Watts.

\appendix

\section*{Appendix A: SWIFTS}
\label{sec:swifts}
The standard autocorrelation is usually measured with a Michelson interferometer and a slow detector, with a scanning mirror inducing a delay $\tau$ between two paths, and the interference measured as a function of this delay.
Such a measurement contains no phase information.
With SWIFTS, the fundamental beat note is instead measured.
As it is the sum of all heterodyne tones between neighbouring modes, a measurement of the beat note will be sensitive to the phase relationship between these modes.
The beat note is demodulated in quadrature as a function of mirror delay:

\begin{equation}
x(\tau) = \left<E(t+\tau)E^*(t)\cos(\omega_r t)\right>
\end{equation}

\begin{equation}
y(\tau) = \left<E(t+\tau)E^*(t)\sin(\omega_r t)\right>
\end{equation}

It can be shown that, for a comb source, and a sufficient total path delay, these quadratures give access to the intermodal phase differences after a Fourier transform.

\begin{equation}
X(\omega)-iY(\omega) = \sum_n |A_n||A_{n-1}|e^{i(\phi_n-\phi_{n-1})}\delta\left[\omega-(\omega_0+n\omega_r)\right]
\end{equation}

This is a direct measurement of the field group delay.
The phase profile is then reconstructed by summing across these phase differences.

For a full treatment of SWIFTS, see for example \cite{burghoff_terahertz_2014,burghoff_evaluating_2015,burghoff_notes_2018}.
The setup used in this paper (marked "Acquisition" in Fig.~\ref{fig:setup}) is described in more detail in the Supplementary of \cite{singleton_evidence_2018}.

\section*{Appendix B: Device characteristic}
\label{sec:devicechar}
\begin{figure}[!htb]
	\centering
	\includegraphics[]{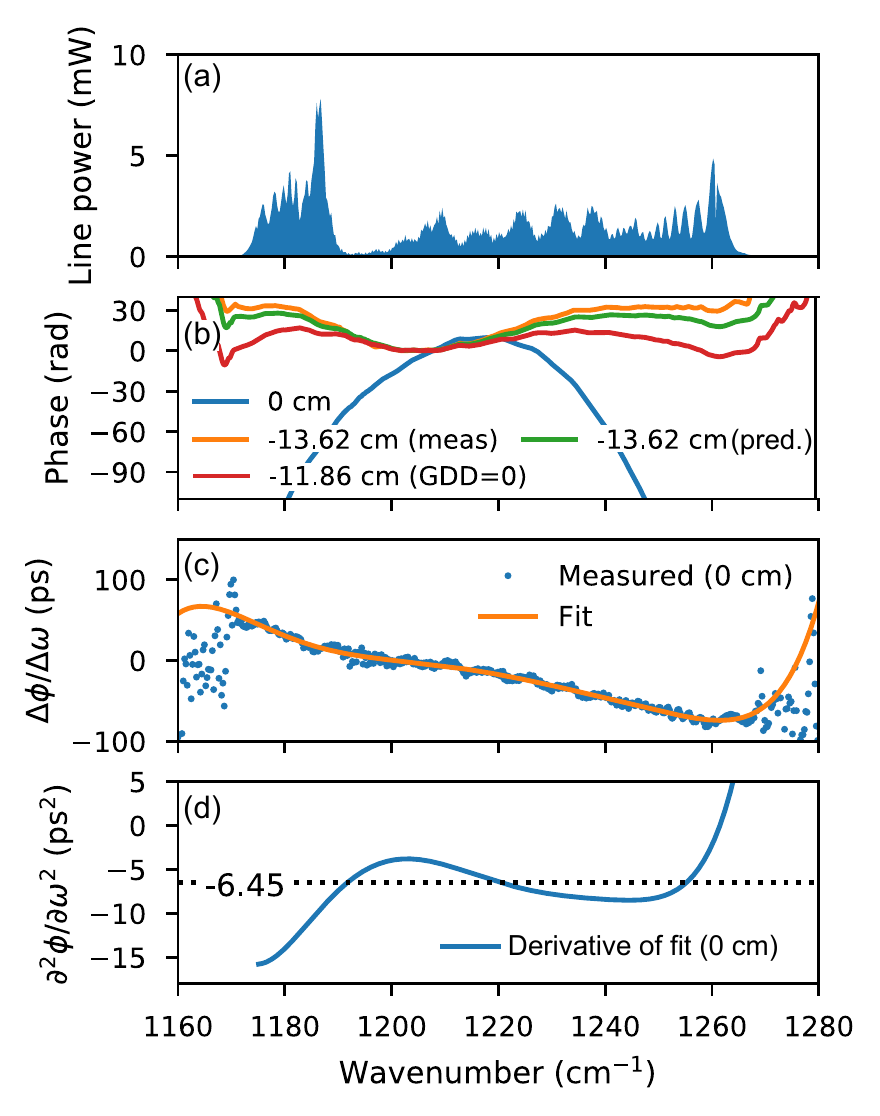}
	\caption{\label{fig:afig_device_char}Device characteristics, measured at 20$^\circ$C, \lasercurrent.
		(a) Optical spectrum, normalised to show the power per mode, as measured  directly in front of the laser.
		(b) Phase spectrum as found by integrating the measured phase difference spectrum in (c).
		(c) Laser output field group delay measured at the nondispersive grating offset 0 cm. The orange curve is an 11$^{th}$ order polynomial fit.
		(d) Field group delay dispersion, calculated as the numerical derivative of the black curve with respect to the circular frequency ($\delta^2 \phi/ \delta \omega^2$)}
\end{figure}

The power spectrum, measured at low-frequency on the MCT detector, and field group delay $\partial \phi / \partial \omega$, measured at the device round-trip frequency of 7.42~GHz, are plotted in Fig.~\ref{fig:afig_device_char} (a) and (c).
The latter is calculated by normalising relative phases between the modes $\phi_{n}-\phi_{n-1}$, extracted from the SWIFTS traces, to the circular cavity round-trip frequency $\omega_r$.
By summing across along this trace, the phase spectrum can be recovered.
This is plotted in (b), where at 0~cm the parabolic profile is very clear.
Also plotted are the predicted and measured phase traces at -13.6~cm, showing good agreement, and the curve with no group delay dispersion for comparison.
An $11^{th}$ order polynomial fit is made to the 0~cm trace and its derivative with respect to $\omega$, that is to say the group-delay dispersion Eqn.~\ref{eq:gdd}, is taken and plotted in (d). Qualitatively, this highlights that the field GDD of the comb possesses third- and higher-order dispersion, which neither the grating compressor, nor any other mainly quadratic compressor e.g. a fibre, can compensate for alone.

\section*{Appendix C: Beat note interferograms}
\begin{figure}[!htb]
	\centering
	\includegraphics[width=\linewidth]{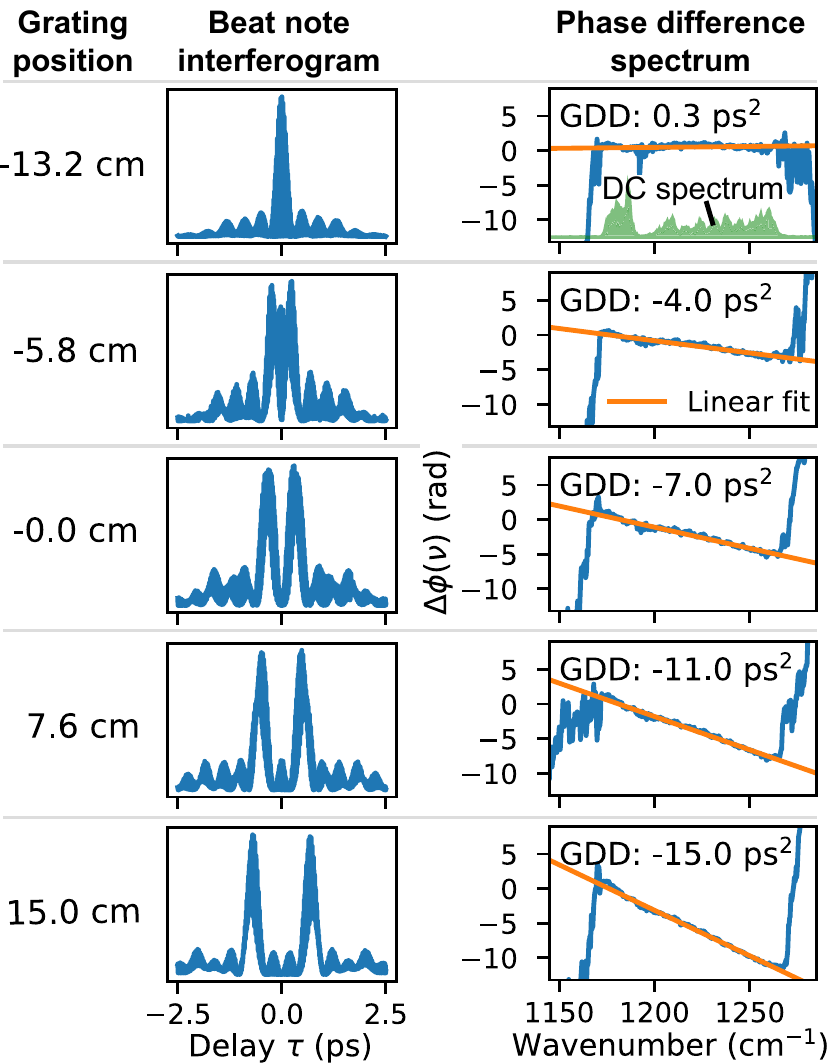}
	\caption{\label{sfig:bn_ifg}Unnormalised beat note interferograms measured at 16~cm$^{-1}$ as a function of the grating displacement (left) with the matching intermodal phase difference trace (right), where 0.0~cm is the reference (unmodified) position, and -13.2~cm represents a strongly AM signal.
		The grating position is noted in the left hand column.
		Fits to the phase difference spectra are plotted in orange, and the estimated GDD noted.
		A green DC spectrum is provided for reference.
	}
\end{figure}

The action of the grating on the field can be seen directly on the beat note interferogram, which shows the RF power in the beat note as a function of the interferometric path difference\cite{hugi_mid-infrared_2012}.
In Fig.~\ref{sfig:bn_ifg}, we plot at five different grating displacements $|x(\tau)+iy(\tau)|^2$, a quantity closely related to the beat note interferogram,, measured at the centreburst with a resolution of 16~cm$^{-1}$.
The reference point is indicated at a displacement of zero, where the grating sits at exactly the focal spot, and the system is dispersionless.
Here, the FM character of the field is demonstrated very clearly, with a minimum at the zero path difference\cite{hugi_mid-infrared_2012,khurgin_coherent_2014}, and a maximum at approximately $1/2B$, where $B$ is the optical bandwidth.
As the second grating is moved closer to the first, the correlation starts to become strongest at the zero path-difference.
The beat note interferogram can be seen to more and more resemble the standard autocorrelation trace, as the spectral components are moved in phase.
In the case of an AM signal, the ratio of the central fringe to that of the maximum would have a value of 1 by definition, since placing the light through an unbalanced interferometer should not lead to an enhancement of the beat note strength.
For more FM signals the opposite is true, with large enhancements expected.

\section*{Appendix D: Timing jitter}
\label{sec:ajitter}
We evaluate the integrated timing jitter by following the procedure of Von der Linde\cite{von_der_linde_characterization_1986,kefelian_rf_2008}:
\begin{equation}
\label{eq:sigt}
\sigma_T = \frac{1}{2\pi m f_r} \sqrt{\int_{f_d}^{f_u} S(f) df}
\end{equation}

and the pulse-to-pulse timing jitter:

\begin{equation}
\label{eq:sigpp}
\sigma_{pp} = \frac{1}{m \pi f_r} \sqrt{\int_0^{+f_r/2} S(f)\sin^2{\left(\frac{\pi f}{m f_r}\right)}}
\end{equation}

where $S(f)$ is the measured single-sided phase noise power spectral density, $m$ is the electronic beat note harmonic index (fundamental $m=1$), and $f_r$ is the fundamental beat frequency.

Beat notes up to the fourth order were acquired at the point of maximum compression (see Fig.~\ref{fig:jitter}, main article), and Gaussian, Lorentz, $1/f^2$ and pseudo-Voigt fits performed on them.
The results are tabulated in Table~\ref{tab:jitter}.
As can be seen, the gaussian FWHM increases approximately linearly with $m$, indicating that the correlation time of the jitter noise is much longer than the pulse repetition time\cite{eliyahu_noise_1996}.
For $\sigma_T$, we evaluate Eq.~\ref{eq:sigt} from the $1/f^2$ fit, and the pulse-to-pulse jitter \ref{eq:sigpp}, which must also include the noise up to 0 Hz, from the pseudo-Voigt. Note that this tends to overestimate the pulse-to-pulse jitter.

\begin{table}
	\centering
	\begin{tabular}{rrrrr}
		\hline
		m &   $\delta_G$ (Hz) &   $2\gamma$ (Hz) &   $\sigma_{T} (fs)$ &   $\sigma_{pp} (fs)$ \\
		\hline
		1 &             399.9 &            334.5 &                  334.6\textsuperscript{\textdagger} &                 11.2 \\
		2 &             571.9 &            440.8 &                  213.0 &                  2.0\textsuperscript{\textdagger} \\
		3 &            1264.2 &            906.6 &                  431.1 &                  2.9 \\
		4 &            1776.4 &           1420.0 &                  400.7 &                  2.9 \\
		\hline
	\end{tabular}
	\caption{Beat note parameters and timing jitter extracted from fits to beat note orders $m=1,2,3,4$.. $\delta_G$ and $2\gamma$ are the FWHM for the Gaussian and Lorentzian fits, respectively. $\sigma_{Tint}$ is the integrated timing jitter, evaluated 20~kHz to 100~MHz, and $\sigma_{pp}$ is the pulse to pulse timing jitter. \textdagger Indicates the value quoted in the main article. }
	\label{tab:jitter}
	
\end{table}

\bigskip
\noindent\textbf{Funding.} Schweizerischer Nationalfonds zur F\"orderung der Wissenschaftlichen Forschung (SNF) (SNF200020-165639).

\bigskip
\noindent\textbf{Acknowledgements.} We thank Dmitry Kazakov for stimulating discussions, and Martin Francki\'e, Mehran Shahmohammadi, and Zhixin Wang for proofreading the manuscript and for constructive feedback.

\bibliography{biblio}

\end{document}